\documentclass{icldt}
\usepackage{amsmath}
\usepackage{graphicx}
\usepackage{wrapfig}
\usepackage{booktabs}

\usepackage{float}
\restylefloat{table}

\title{The Role of Social Feedback in Financing of Technology Ventures}
\author{Aleksandar Bradic}
\date{September 28, 2012}
\department{SOAS}
\supervisor{}
\dedication{I would like to express sincere gratitude to my thesis supervisor, Norman Williams, for his help in preparation and production of this work. Special thanks also goes to Kevin Laws, my long-standing business mentor and person that had introduced me to the fascinating world of Venture Capital Economics. Finally,  deepest gratitude goes to my loving wife, Jelena Bradic for all of her support and encouragement throughout this process.}

\begin{document}

\maketitle

\begin{abstract}
This research examines relationship between staging of Venture Capital (VC) investments and "social feedback" visible in publicly available data on the Web. We address the question of Venture Capital investment sensitivity to performance and prospects of new venture, given as likelihood of obtaining future financing, available exit options and duration between investment rounds. We argue that in the case of Internet companies, publicly available "social feedback" data, such as search trends and website traffic information, can be used as a proxy for some of company's internal metrics such as user base growth and product adoption. In order to answer questions of interest, we compile unique dataset consisting of detailed information about Venture Capital investments in the Internet Technology sector over the period from 2004 to 2012 and associated longitudinal search trend and website traffic data. By applying methods of survival analysis, we find that positive trends in search and website traffic volumes can lead to increased likelihood of future financing and shortening of duration between subsequent financing rounds. We also find evidence that Òsocial feedbackÓ only impacts company's ability to attract next round of financing or exit via IPO, while M\&A exits seem relatively independent of such performance metrics and can occur at any stage of company development. Such findings provide strong evidence in support of Òlearning hypothesisÓ and suggest VC's ability to identify prospects of new venture early in it's development and allocate funding accordingly. Given research also provides methodological contributions to the problem of evaluating the prospects of new startup companies using only publicly available data, and as such should be of interest in applications such as new investment screening and industry-level assessments by analysts or policy makers.
\end{abstract}

\makededication
    
\tableofcontents

\chapter{Introduction}

Venture Capital represents dominant way of obtaining financing for new technology ventures. It is estimated that, only in the first quarter of 2012, Venture Capitalists have invested \$5.8 billion across 758 deals, out of which 41.6\% in the areas of Software, IT Services, Media and Entertainment \textit{(PriceWaterhouseCoopers MoneyTree Report, April 2012)}. Venture Capitalists usually employ Òstage financingÓ by investing in portfolio of companies across multiple rounds, between which progress and potential of each venture is evaluated and decisions are made regarding future financing and preferred exit routes. In addition to addressing potential moral hazard and related issues, staging of VC financing provides Venture Capitalists with an opportunity to reevaluate the potential of new venture and select exit route that maximizes expected return on investment. These decisions are primary based on company's internal metrics such as user adoption, rate of growth, intellectual property, cash flow, but also depend on external factors such as industry trends and market conditions.

The subject of Venture Capitalist's decisions making regarding staging of investments has been an active area of research, focusing on questions such as factors affecting investment decisions, their size and duration. In most of the studies, researchers focus either on the analysis of external factors such as VC characteristics, previous financing and market conditions or firm-specific factors such as financial capital, cash flow, intellectual property and firm structure. However, the main challenge regarding analysis of the impact of firm-specific factors is that most of this information is not publicly accessible and therefore only available for companies that have gone through the IPO process. This leaves a significant gap in terms of analysis of the impact of firm-specific factors on investment prospects for privately held companies by parties with access limited to publicly available data. Such analysis could be beneficial to wide variety of applications, ranging from Venture Capitalist investment screening for new later-stage investments to evaluation of prospects of competitor companies. Perhaps the most important application of such result would be industry-level assessments or forecasts, by analysts or policy makers, which tend to be inherently limited to publicly available data.

In this research, we argue that in case of Internet Companies, publicly available Òsocial feedbackÓ data, such as search trends and website traffic information, can represent a reasonable proxy for some of the company's internal metrics, such as growth and user adoption. This data has become increasingly available over the course of last decade and provides the level of transparency and real-time insight into development of new technology companies in a way that was previously unknown. Therefore, we hope that given data should be instrumental in gaining better understanding of Venture Capitalist decision-making process regarding financing and exit decisions in startup companies. We should also note that, over the last decade, Internet companies have become a significant part of most Venture CapitalistÕs investment portfolios, while relatively little academic research has been conducted focusing specifically on VC activity in Internet Technology sector following the period after dot-com boom. In given paper, we aim at filling this gap by compiling a unique dataset consisting of detailed information about majority of VC investments in Internet (Òdot-comÓ) companies, over the period between 2004 and 2012, along with corresponding search trend and website traffic data and provide detailed analysis of investment round sizes, duration and exit options of portfolio companies.

The main question that we address as part of this research is the analysis of VC decision-making regarding follow-up investments in new technology companies in the light of publicly available Òsocial feedbackÓ data. In particular, we aim at testing the hypothesis of VC rationality, active monitoring of investments and ability to evaluate prospects of new ventures at early investment stages by formulating the following hypothesis:\\

\textit{H1. Positive trends in ÒSocial FeedbackÓ data are expected to result in increased likelihood of obtaining next round of financing for VC-funded technology companies.}\\

Given hypothesis suggests that companies which exhibit positive trends in social feedback data are more likely to be perceived by VCs as having a higher growth potential and represent better candidates for future investments. Given that new startup companies face a limited window of opportunity in which they can develop new product and capture significant market share, it is likely that VCs might have an incentive to invest in such companies more aggressively, shortening the duration between subsequent investment rounds. We aim at testing this aspect of VC investment by formulating the following hypothesis:\\

\textit{H2. Positive trends in ÒSocial FeedbackÓ data are expected to result in decrease of time duration between two subsequent rounds of financing for VC-funded technology companies.}\\

Finally, we note that each venture faces a number of possible outcomes at each stage in its development: receiving a next round of financing, exit via IPO or M\&A or termination due to the lack of funding or availability of other exit options. While we expect company's performance metrics to be highly indicative of its likelihood of obtaining future financing, termination or IPO exit, we expect these to have much weaker influence of likelihood of M\&A exits. This should be particularly expected in the case of Internet companies, which have witnessed a large number of Òacqui-hireÓ exits over the course of last decade, in which startups get acquired at relatively low value mostly as a mean of acquiring high-profile employees and potential intellectual property assets, with little regard for actual business performance of acquired companies. Such exist might also be indicative of VC's ability to orchestrate "soft" exist in the cases where portfolio company manages to build a great team or develop valuable technology, but fails to generate viable business around it. Such exits could potentially enable VCs to recover some of assets invested in portfolio company and have a minor success story, even if expected multiple-digit returns from successful exits have not been materialized. In order to address this aspect of VC investments, we formulate the final hypothesis:\\

\textit{H3. Likelihood of M\&A exits for VC-funded technology companies is not expected to be significantly determined by trends in ÒSocial FeedbackÓ data.}\\\\

The rest of the research is organized as follows. In Section 2 we provide a detailed overview of existing literature on the subject of VC decision making regarding staging of capital infusions in portfolio companies and its determinants, as well as the literature on various types of "social feedback" data and its applications to modeling of real-world economic variables. In Section 3, we introduce formal procedures and methods we will be using in order to conduct the research. We outline the data collection methodology in Section 4 and provide a detailed description of the obtained dataset schema in Section 5. We perform an analysis of given dataset in Section 6 and present final results in Section 7 along with corresponding implications to proposed set of hypothesis and main research questions. Finally, we conclude with brief summary of key findings and its potential implications, as well as open questions and motivation for future work.

\chapter{Literature Review}

Venture Capital plays a critical role in innovation cycle by providing financing for early stage, high potential, high risk, growth startup companies. Such companies find obtaining financing through traditional mechanisms virtually impossible due to the four critical factors: uncertainty, asymmetric information, the nature of firm assets and conditions in the relevant financing and product markets \cite{GL06}. Venture Capitalists aim at addressing some of these issues by playing the role of informed screening agents, providing superior evaluations of project quality, taking active role in company development and monitoring company's prospects and performance \cite{MY10}. Upon an actual financial investment in the firm, in addition to providing capital and advice, VCs either grant the investee firm access to their existing network of contacts across technology experts, intellectual property consultants, suppliers, purchasers, investment banks and legal and accounting advisors or help the firm cultivate such a network \cite{CJ10}. VCs also tend to play a critical role in facilitating complex networks of innovation in a way that enables participating parties to gain competitive advantage and increase likelihood of projectÕs success \cite{FG09}. Presence of VCs also tends to reduce time to market for new products \cite{HP00} and help in conveying credible signal of firmÕs quality to third parties \cite{CEA10}.

A particularly important research question considering Venture Capitalists investment process is gaining better understanding of VC's tendency to stage investments by spreading them across multiple rounds over time and decision-making process regarding future investments at each milestone in company's development. Predominant view in existing literature is that the main reason for VCs to stage their investments is control of risk and mitigation of agency problems \cite{WZ02}. Seminal empirical study in \cite{G95} shows that in financing of high-risk companies with pervasive moral hazards, staged financing allows VCÕs to mitigate some of associated agency problems by having the ability to gather information and monitor the progress while maintaining the option to abandon project at any investment stage.

Given that staging of investments can also have a number of negative results, such as negotiation and contracting costs at each round or inducing the entrepreneur with an aim of short-term rather than long-term success, in \cite{T11} author argues that VC investors tend to balance the cost of staging and effective monitoring of entrepreneur and engage in staging only if effective monitoring of the entrepreneur is too costly. This point of view is also known as "monitoring hypothesis". Alternative explanation of VC's propensity to stage investments is given by "hold-up hypothesis", suggesting that staging of financing can help in mitigating hold-up problems by the entrepreneur \cite{N99}, given that it limits the amount of VC's investment in the venture and therefore reduces the entrepreneur's incentive to leave the firm at any given time. Finally, in \cite{BH98}, authors introduce what is known as a "learning hypothesis", which suggests that staging allows for optimal contract providing intertemporal risk-sharing between venture capitalist and entrepreneur in scenarios where the value of the project is initially uncertain and more information arrives by developing the project. In this context, staging creates value for Venture Capitalist, since it generates a real option for VC to revise project financing and entrepreneur's share at each financing round, depending on information learned between rounds regarding the venture or the entrepreneur. Such position is likely to be of particular interest in the case of Internet companies, which primary operate in breakthrough markets, where evaluation of company's prospect prior to actual product development and initial launch is virtually impossible.

In addition to reasons for staging of VC investments, factors impacting investment structure, timing and size of financing rounds have been extensively studied. In \cite{GL06} authors outline that primary factors influencing investment decisions include company's growth, age, investment volume and industry conditions. Given that the quality of a venture is often not directly observable, VCs tend to base their decisions on a number of additional observable characteristics that might serve as "signals" for evaluation of prospects of young companies. In \cite{H09}, authors suggest that one such signal might be existence and quality of patents filled and held by young ventures. Given that patents can help companies appropriate returns from investments in R\&D and facilitate commercialization of technology, it is expected that such signals should influence VC's decision making. The authors provide evidence that having filled at least one patent application reduces the time to first VC investment by 76\%, as well as that investors seem to be well capable in interpreting such signals to the point of being able to accurately predict the quality of patents measured by number of citations given patents are to receive in the future. Likewise, in \cite{BEA11} authors analyze dependency of VC investments in European Union and appropriate measure of trust between nations and show that trust can have a significant positive effect on likelihood of investments. Similarly, in \cite{ZH07} authors identify four symbolic actions performed by entrepreneurs that can lead to increased frequency and quantity of investments: conveying the entrepreneurÕs personal credibility, professional organizing, organizational achievement and quality of stakeholder relationships. In \cite{HP00}, authors also show that product market size and degree of innovation (innovation vs. imitation) tend to directly influence likelihood of attracting VC investments. Signals used in evaluation don't necessary have to be firm specific. For example, in \cite{GEA07} authors show that VCs tend to react to favorable public market signals, such as the increase of IPO offering valuations, by increasing their investment in entrepreneurial firms. In \cite{JW00} authors show that similar effects can also occur as a result of changes in government policies. In this research we argue that, at least in the case of consumer technology companies, predominant factors influencing VC decisions regarding future investments should also be related to actual adoption of new projects and customer base growth, which should ÒsignalÓ venture's prospects much stronger than a number of above mentioned, less tangible assets. To the best of our knowledge, this subject has not been addressed in current body of research, most likely due to the inaccessibility of data that would allow for such relation to be established.

Another important aspect of Venture Capital cycle that has been studied in the literature is the duration of investments. In \cite{CJ10} authors formulate a theory of VC investment duration based on the idea that venture capitalists exit when expected marginal cost of maintaining the investment becomes greater than expected marginal benefit. Most important exit options available to investors include IPOs, trade sales and liquidations, but a number of other exit options exist, such as share repurchase by the founder or selling shares to institutional investors. Given that preferred exit options of VCs and entrepreneur tend to diverge over time, it is important for companies to have an efficient way of selecting between different exit options (especially between IPO and trade sales). In \cite{BW01} authors show that convertible securities allow for implementation of such ex-ante agreed optimal exit policies, suggesting an explanation for their widespread usage in VC finance. A number of publications also indicate that likelihood of different exit options for VC-funded companies generally has different dynamics of change over time. In \cite{GS05} authors show that likelihood of exit through IPO tends to increase with time, until it reaches a plateau (usually up to four years since initial investment) and then sharply decreases, while likelihood of exits via trade sale tends to vary much less with time and therefore provides a much more ÒuniversalÓ exit option. In \cite{GA10} authors argue that VCs tend to adopt a preferred exit strategy (IPO or acquisition) very early in investment process and engage in resource base adjustments to prepare the firm for respective exit. However, it is also possible for investors to adopt a different exit strategy over time as a result of failure of preferred option (for example, investors might pursue acquisition as exit strategy as a result of company's failure to become public). Additionally, different classes of VCs might have different exit and investment duration preferences - in \cite{GEA11} authors show that corporate VC funds generally tend to allow for longer duration of investments and higher likelihood of exit through acquisition, whereas independent VC funds target shorter duration and larger investment in order to increase the likelihood of exit through an IPO.

Investment duration is likely to be influenced by a number of factors. In \cite{CM01} authors find that such factors include stage of firm at first investment, capital available to the VC industry as whole and whether the exit was preplanned and/or made in response to unsolicited offer. Similarly, in \cite{CJ10}, authors show that investment duration seems to be longer for early-stage and high-tech investments and shorter for investments in entrepreneurial firms that are older at the time of first VC investment. Authors also find that investment durations generally tend to be shorter during periods of strong market conditions as well as in the case of syndicated and larger investments. In \cite{GS05} it is shown that firm industry type can have significant impact on investment duration - Biotech and Internet have fastest IPO exits, but while Internet firms are fastest to liquidate, Biotech seem to be the slowest. Authors also show that geographical location of entrepreneurial firm can have a significant impact on the likelihood of trade sales but not on the likelihood of IPO exit. In \cite{SEA08} authors find strong support for the signaling effect, implying that VCs have ability to identify non-performing investments and tend to write-off such investments instead of continuing to commit further capital. Authors also find that positive market sentiment and generally favorable stock market climate tends to increase probability for a buyout backed IPO exit.

In this research, we argue that actual observed company's performance is likely to have a significant impact to investment duration, given that most of start-ups only face a limited window of opportunity for new product development and marketing, and therefore VCs that identify projects with higher chance of success are expected to increase frequency and shorten duration between new investments. In literature on VC exits and investment duration, several publications have particularly focused on factors influencing survival of Internet firms. In \cite{KW07} authors show that market, firm and e-commerce related variables, such as the entry of additional Internet firms via IPOs, a smaller firm size, IPO timing, late entry and selling of digital products or services can reduce Internet firm's likelihood of successful exit. Authors also find that Internet firms operating in breakthrough markets (such as online portals or auction sites) are more likely to survive due to less competitive pressure from traditional businesses. In \cite{CW07} authors show that presence of technology-related patents can serve as a signal of firm's quality and increase likelihood of Internet company survival. Study in \cite{C04} also finds that factors such as reputation of participating VC firms, total amount of investments raised and size of startupÕs network of strategic alliances can reduce time and increase likelihood of startupÕs exit via IPO. In \cite{BEA07} authors also find that relationships between factors influencing Internet firm's survival may vary over calendar time - in early investment stages, firm's survival is generally closely related to the IPO rate of Internet stocks and abundance of financing capital, whereas in later stages, survival tends to be more associated with firmÕs financial capital and size.

In this research, we argue that Òsocial feedbackÓ information, expressed as implied demand for given project visible in search and social media websites, should also represent a significant factor influencing survival of Internet companies. This information fits pretty well the definition of "signal" given as "characteristic that is correlated with company performance, but easier to observe than underlying causal factor influencing performance" \cite{H09}, due to itÕs direct correlation with certain internal metrics, such as customer base growth and product adoption. Given its public nature, this information can also perform a role of product quality signaling and allow VCs to directly monitor progress of the project, reducing information asymmetries and associated agency issues. In this context, we can hypothesize that positive trends regarding entrepreneurial project visible in public data should reduce investorÕs uncertainty and increase its likelihood of participation in subsequent investment rounds. To the best of our knowledge, such role of Òsocial feedbackÓ in financing of Internet companies has not yet been investigated in literature.

The main reason why this information is expected to be indicative of technology ventureÕs success is reflected in the way Internet and Social Media have changed the means by which consumers learn about and adopt new products. In \cite{W10} author outlines three broad categories of Internet-based media that have most significantly changed the way in which consumer attention takes shape: \textit{Search Engines} (such as Google or Yahoo), \textit{Content Providers} (online versions of traditional media like New York Times or user-generated content like Wikipedia and YouTube) and \textit{Social Networks} (such as Facebook, MySpace, Digg and others). All of these entities have one thing in common, which is that compared to traditional media, they offer some measure of interactivity and allow users to provide active feedback by voting, sorting, retrieval, recommendation, commenting and sharing their opinion with fellow users. It is estimated that over 75\% of Internet currently users use ÒSocial MediaÓ by joining social networks, reading blogs or contributing user reviews \cite{KH10}. A lot of user activity in Internet media seems to be related to brands or products. A study in \cite{JEA09}, on the usage of microblogging, shows that 19\% of all microblogs contain mention of the brand, out which 20\% contains some expression of sentiment, with 50\% being positive and 33\% being critical of company or product. In \cite{HTEA12} authors show that microblogging word of mouth through Twitter and similar services can significantly influence the success of new products by shifting early adoption behaviors. Usage of Internet-based media can result in social contagion, which is likely to impact affect new product diffusion and itÕs adoption among consumers \cite{LEA12}. In \cite{PEA10} authors formalize the notion of diffusion processes of new products and services using the concept of "Innovation Diffusion", given as "the process of the market penetration of new products and services that is driven by social influence, which include all interdependencies among consumers that affect various market players with or without their explicit knowledge". In \cite{CHEA10}, authors formulate a model of diffusion of digital/information products based on number of competitors and characteristics related to innovation and product bundling. In \cite{CH11} author investigates the role of online buzz in new product diffusion and finds that it can both accelerate the processes of new product diffusion by influencing imitation tendency and expand itÕs potential market size.

Given that most of consumer activity in Internet media results in publicly available "social feedback" data, a number of publications have dealt with the problem of using this data in predicting various aspects of product or company's success. For example, in \cite{JW08} authors show that online consumer reviews and ratings can have significant impact on sales, prices and profits. Authors outline a model that explains interplay between consumer ratings and informativeness of the reviews and point out conditions under which consumer rating improvements can either benefit or hurt firmsÕ sales and profits. Similarly, in \cite{DUEA08} authors analyze online user reviews as endogenous rather than exogenous factors related to movie box office sales and find that, even though ratings of reviews in this context don't have significant impact on box office sales, the volume of online reviews does have a significant impact, suggesting the ability of online reviews to accurately reflect consumer awareness. In \cite{GREA05} authors pose the general question of predictive power of online chatter in the form of blogs, bulletin boards, web pages, wikis and related collaborative technologies and show that online postings can successfully predict spikes in sales ranks. Similarly, research in \cite{TT12} examines whether aggregated user-generated content (UGC) from a number of websites can be related to stock market performance and find that volume of chatter can have a strong positive effect on abnormal returns and trading volume, leading the abnormal returns by a few days (supported by Granger causality tests). In \cite{BOEA10} authors analyze sentiment of text content in Twitter feeds and find that Òcollective moodÓ defined in this manner can be used to significantly improve predictions of daily changes in Dow Jones Industrial Average closing values. Recently, a number of publications have also addressed the question of predictive power of social media regarding election results. For example, in \cite{TUEA10} authors show that mere number of messages mentioning a party accurately reflects the election results as well as that tweetÕs political sentiment tends to closely correspond to parties political opinions, indicating Twitter message's ability to accurately reflect offline political landscape. However, as \cite{GAEA11} suggests, long-term predictive power of social media such as Twitter regarding electoral results can be quite limited duo to inherent biases present in the data as well as itÕs propensity for manipulation by spammers and propagandists.

Web Search also represents a particularly important aspect of Internet-based media, given its role as primary mean of information discovery for most of Internet users. Therefore, it should come as no surprise that trends in Web search queries are capable of accurately predicting various aspects of consumer behavior. In \cite{CV09A} authors show that search engine query data can be used to forecast near-term values of economic indicators, such as monthly automobile retail sales and in \cite{CV09B} that it can be also used to predict additional indicators such as initial claims for unemployment. In \cite{AZ09} authors further deal with question of unemployment indicators and find strong correlations between Google keyword searches and unemployment rates, indicating that search trends might provide a ÒcontinuousÓ indicator of certain economic variables which are otherwise reported only periodically. In \cite{DAM10}, authors take this notion a step further and construct a ÒGoogle Job Search IndexÓ based on Google search data and show that models augmented with such index perform significantly better than traditional ones in predicting US unemployment rates. Similar index, called Google Inflation Search Index (GISI) is constructed in \cite{GU10} and shown to provide accurate indicator of inflation expectations with lowest forecast error of all tested expectation indicators. A number of publications also suggest potential of search query data in forecasting additional economic indicators such as Private Consumption \cite{SV09}, Housing Prices \cite{MR11} and Consumer Sentiment \cite{PH09}. In addition to macroeconomic trends, search query data can be very effective in predicting public demand regarding particular subjects. For example, in \cite{DAEA11} authors show that aggregate Google search frequency for individual stock symbols from Russell 3000 index seem to be strongly correlated with but different from existing proxies of investor attention. In \cite{GOEA10} authors find that search volumes can be highly predictive in forecasting opening weekend box-office revenues for feature films, first-month sales of video games and ranks of songs on Billboard Hot 100 chart. Finally, in \cite{BTS11}, authors use Google Trends to identify public interest in different science-related topics, in \cite{FIG11} to characterize growth patterns of YouTube video popularity and in \cite{REC07} in discovering trends in software engineering. Interestingly, to the best of our knowledge, no publications have dealt with the question of using search query data in order to determine consumer interests in Internet companies and related products.

In this research, we aim at addressing the question of impact of "social feedback" on survival and financing of Internet companies. Given the challenges associated with attempting to aggregate all of data across all possible Internet media of interest, we restrict ourselves to the analysis of two most important indicators - Search Query Volume and Website Traffic. Website Traffic is especially important indicator as it represents a direct measure of product market performance in the case of Internet firms \cite{DL03}, \cite{HEI02}. Given the nature of information diffusion on the Internet, we expect that between the two variables we should be able to capture a significant signal of consumer demand for particular companies and products. We expect this research to contribute to existing body of knowledge in several ways. We aim at providing empirical support for "learning hypothesis" regarding staging of VC investments, in the light of publicly available performance-related indicators, given in the form of Web Search Trends and Website Traffic, and analyze itÕs impact to various aspects of investments in technology companies such as likelihood of attracting next financing round, duration between investments and exit options. We also attempt at addressing potential differences in VC decision making regarding exit options of technology start-ups and its sensitivity to companyÕs performance. Finally, we expect to provide an example of usability of "social feedback" data in analysis of companyÕs prospects and provide a methodological contribution to the body of research on technology start-up prospect evaluation in applications limited to publicly available data. To the best of our knowledge, no such research has been conducted in existing literature.

\chapter{Procedure and Methods}

In order to address research question of interest, we formulate the problem as estimation of the likelihood of obtaining next round of financing and duration between each two financing rounds as a function of "social feedback" information. Such problem has a convenient representation in the framework of survival analysis, with next round of financing or liquidity events (M\&A, IPO) representing events of interest and duration of time between subsequent financing rounds representing survival times. In the same context, entries for which events of interest did not occur for the period of analysis are considered Òright-censoredÓ. The framework of survival analysis represents a ÒnaturalÓ context for analysis of such censored duration data and should be instrumental in answering main research questions of this work. This should be particularly the case given that survival analysis methodology has already been successfully used in literature in addressing various aspects of VC investment process and performance of start-up companies (for example - \cite{CW07} , \cite{KW07}). In this chapter, we introduce main concepts of survival analysis, outline proposed model that we will apply to the research questions at hand and describe methodologies that we will be using in addressing the validity of results and hypothesis testing.

Assuming that $f(t)$ represents probability density function of time-to-event $T$, we define \textit{survival function} as probability of financing event not occurring before time $t$:

\begin{center}
$S(t) = Pr(T > t) = \int_0^t \! f(x) \, \mathrm{d} x. $
\end{center}

We define \textit{hazard function} as instantaneous rate of events at time $T=t$, given that event has not occurred up to time $t$ as:

\begin{center}
$h(t) = \lim\limits_{\Delta_t \rightarrow 0+}{P(t < T \leq t + \Delta t | T > t) \over \Delta t} = {f(t) \over S(t)}$
\end{center}

In this context - $h(t)\Delta t$ represents approximately probability of occurrence of event of interest in the $(t, t + \Delta t]$ interval, given that the event has not occurred up to time $t$. Therefore, in our context, hazard function $h(t)$ can be interpreted as a likelihood that company will receive next round of funding (or exit via merger or IPO) in each period.

In this research, we're particularly interested in modeling hazard function of Venture Capital investment duration as a function of appropriate explanatory variables, including "social feedback" data. In order to do so, we use semi-parametric model known as Cox proportional hazard regression model. Basic Cox model with fixed covariates is defined as:

\begin{center}
$h_i(t) = h_0(t)e^{(\beta_1x_{i1}+...+\beta_kx_{ik})}$
\end{center}

where hazard for company $i$ at time $t$ is given as the product of two factors - $x$ representing vector of explanatory variables and $h_0(t)$ representing unspecified baseline hazard function. Baseline hazard function is generally interpreted as hazard function that is "common" for all companies and corresponds to hazard function of entry for which values of all covariates have the value of zero.

Cox proportional hazard model is defined in relative, rather than absolute terms and can be represented as a linear function of logarithm of firm-specific and baseline hazard ratios:

\begin{center}
$log\{{h_i(t) \over h_0(t)}\} = \beta_1x_{i1} + ... + \beta_kx_{ik}$
\end{center}

Given the nature of Venture Capital investment process (limited partnership structure of VC funds, investor "learning" throughout the lifecycle of the venture) and empirical observations about duration and exit options of VC investments \cite{GS05}, it seems reasonable to assume that, in addition to firm-specific factors, all of VC-backed companies in a single industry face a set of common investment risks. Such common risks can be captured by given baseline hazard function $h_0(t)$. Therefore, Cox proportional hazard seems appropriate as it enables us to decompose hazard function into set of firm-specific factors and baseline hazard, common to all companies.

In order to estimate Cox model, we construct appropriate log-likelihood function given as:

\begin{center}
$L(\beta) = \sum_{i=1}^{n}\{c_i ln[h_0(t_i)] + c_ix_i\beta + e^{x_i\beta}ln[S_0(t_i)]\}$
\end{center}

where $c_i$ represents value of censoring variable at observation $i$ and $S_0(t)$ baseline survival function given as $S_0(t) = e^{-H_0(t)}$ with $H_0(t) = \int_0^t h_0(u) du$ representing cumulative baseline risk.

However, given that in the case of Cox model baseline hazard and survival functions are not specified, it is not possible to obtain estimates of parameters of interest $\beta$ by simple maximization of given full likelihood function. Instead, it can be shown that maximization of properly defined "partial likelihood function" dependent only of parameters of interest can yield parameter estimators with the same distribution properties as full maximum likelihood estimators. Such partial likelihood function for Cox model is defined as:

\begin{center}
$l_p(\beta) = \prod_{i=1}^{n}[{{e^{x_i\beta}} \over {\sum_{j\in R(t_i)}{e^{x_j\beta}}}}]^{c_i}$
\end{center}

where summation in the denominator is over all subjects at risk at time $t$ given as $R(t_i)$. Given function is modified to exclude right-censored events for which $c_i=0$, and defined over total of $m$ ordered survival times and final log partial likelihood function defined in this manner is given as:

\begin{center}
$L_p(\beta) = \sum_{i=1}^{m}\{x_{(i)}\beta - ln[\sum_{j\in R(t_{(i)})} e^{x_j\beta}]\}$
\end{center}

where $x_{(t)}$ denotes the value of covariate for the entry with ordered survival time $t_{(i)}$. By maximizing given log partial likelihood function, we obtain maximum partial likelihood estimator $\beta$ of Cox proportional hazard model.

It is important to note that given model depends on the fact that we can leave baseline hazard unspecified. This implicitly contains two assumptions: multiplicative relationship between underlying hazard rates and log-linear function of covariates (proportionality assumption) and log-linear effect of the covariates on the hazard function. Assuming that given assumptions are satisfied, estimates obtained in this way are consistent and asymptotically normally distributed. Due to the fact that hazard for any particular entry is fixed proportion of the baseline hazard, hazard ratio for any two entries $i$ and $j$ is given as:

\begin{center}
${{h_i(t)} \over {h_j(t)}} = e^{\beta_1(x_{i1}-x_{j1}) + ... + \beta_k(x_{i1}-x_{j1})}$
\end{center}

representing constant proportion in hazard rates between two entries, which is the main reason that Cox model is referred to as a proportional hazard model.

Given that Cox proportional hazard maximizes partial likelihood function considering only companies for which event has occurred during the period of interest, obtained estimates of coefficients $\beta$ can be interpreted as likelihood of receiving financing relative to all other companies. Percent change in the hazard as result of individual parameter $j$ can be easily calculated from the exponentiated coefficients, as $(e^{\beta_j}-1)*100$. Therefore, $\beta_j$ estimates with values greater than 1 can be interpreted as indication that given covariate $j$ is associated with increased hazard of having the event of interest, whereas estimates with values less than 1 can be associated with decreased hazard of having the event of interest. Finally $\beta_j$ estimates with value of 1 can be interpreted as indicative of no association between covariate and the hazard. We should keep in mind that estimated coefficients correspond to ratio and should be interpreted as odds rather than probability. Therefore, in the case of dichotomous covariates, estimated value $\beta_j$ should be interpreted as the fact that the odds of company having a single value of given covariate receiving financing are $((e^{\beta_j}-1)*100):1$ relative to the company having alternative value of covariate, all other things being equal. Similarly, hazard ratios for fixed-continuous covariates should be interpreted as the amount of change in the hazard of the event for each unit change in the covariate.

In order to address the validity of obtained results and perform hypothesis testing, we also need to introduce a set of appropriate statistical procedures. We note that the standard errors of $\beta$ obtained using maximum partial likelihood are asymptotically normally distributed:

\begin{center}
$\hat{\beta} \sim N(\beta, E\{I(\beta)\}^{-1})$
\end{center}

where $I(\beta)$ represents \textit{observed information}, defined as second derivative of log partial likelihood. This enables us to test for individual parameter significance by simply constructing appropriate confidence intervals and determine corresponding z- and p- values.

In order to test for overall model significance, most commonly used tests are partial likelihood-ratio and Wald test. Partial likelihood ratio test is based on calculating the statistics given as difference between log partial likelihood of the model containing covariates and the same likelihood model without covariates:

\begin{center}
$G = 2\{L_p(\hat{\beta}) - L_p(0)\}$
\end{center}

where $L_p(0) = -\sum_{i=1}^{m} ln(n_i)$ and $n_i$ represents number of subjects in the risk set at observed survival time $t_{(i)}$. Under the null hypothesis that all coefficients are equal to zero, given statistics follows chi square distribution with $k$ degrees of freedom, which we can use to obtain appropriate p-values and test for the significance of the model. Similarly, Wald test is based on statistics defined as:

\begin{center}
$\hat{\beta}^T[I(\hat{\beta})]\hat{\beta}$
\end{center}

which, under the null hypothesis that all coefficients are equal to zero, follows a chi square distribution with $k$ degrees of freedom, enabling us to construct appropriate confidence intervals and obtain corresponding p-values. In practice, likelihood-ratio test is generally preferred over the Wald test as a way to assess overall model significance. In our analysis, we compute both Wald and likelihood ration statistics for each estimated model and use obtained p-values in order to assess overall significance of the model. We report values of both statistics and test for appropriate critical values.

In addition to estimating goodness of fit of the model and significance of individual parameters, it is important to test for potential violation of proportional hazard assumptions. Proportional hazard model assumes that the hazard for any particular subject is a fixed proportion of any other subject over time. Violation of given assumption might read to incorrect estimation of relative risk Ð if a covariate has hazard ratio that increases over time, relative risk will be overestimated, while in the cases when hazard ratio decreases over time, relative risk will be underestimated. Additionally, violation of proportional hazard assumptions can lead to incorrect estimation of standard errors of parameter estimates, which will result in decrease of significance test power. In order to test for violation of proportional hazard assumptions, we use the method of Schoenfeld residuals, representing difference between observed and expected values of the covariates at each failure time. The $k$th Schoenfeld residual, defined for entry $k$ on $j$th explanatory variable is given by:

\begin{center}
$r_{s_{jk}} = c_k\{x_{k}^{(j)} - a_{k}^{(j)}\}$
\end{center}

where $c_k$ represents $k$th entry's censoring indicator, $x_{k}^{(j)}$ the value of $j$th explanatory variable on $k$th entry and $a_{k}^{(j)}$ is given by:

\begin{center}
$a_{k}^{(j)} = { {\sum_{m\in R(y_k)} x_{m}^{(j)} e^{ x_{m}^{'} \hat{\beta} }} \over { \sum_{m \in R(y_k)} e^{x_{m}^{'} \hat{\beta} } } }$
\end{center}

and $R(y_k)$ represents risk set at time $y_k$. Schoenfeld residuals have the property that, in large samples, expected value of each residual $r_{s_{jk}}$ is zero and that residuals are uncorrelated with each other. If assumptions of proportional hazard hold, none of obtained residual values should be time-dependent. Therefore, a simple graphical method for testing of proportional hazard violations would be plotting obtained values of Schoenfeld residuals against time and checking whether a particular coefficient from a covariate is time-dependent. Alternatively, we can use a test procedure by (Grambsch and Therneau, 1994), which is based on Schoenfeld residuals scaled by an estimator of its variance:

\begin{center}
$\hat{r_{l}^{*}} = [ \hat{Var} ( \hat{r_l} )]^{-1} \hat{r_l} $
\end{center}

where $\hat{Var} ( \hat{r_l} )$ represents estimator of the $p * p$ covariance matrix of the vector of residuals for $i$th entry, with residual values omitted for right-censored entries. In this test, we consider time-varying coefficients $\underline{\beta}(t) = \underline{\beta} + \underline{\theta}g(t)$, where $g(t)$ is a predictable process and test for $H_0 : \underline{\theta} = 0$. Since it can be shown that scaled Schoenfeld residuals $\hat{r_{l}^{*}}$ have approximately mean of $\underline{\theta}g(t_k)$, it is possible to derive a generalized least-squares estimator of the coefficients and a score test of the hypothesis that given values are equal zero assuming specific choice of the function $g(t)$. In our analysis, we perform Grambsch-Therneau test for each estimated model and use obtained $p$-values to test for null hypothesis of no time dependence of coefficients. In cases when given hypothesis is rejected, we conclude that given variable violates proportional hazard assumptions. In order to accommodate such variables, we introduce interaction effect by building interactions between given variable and time in the regression model.

Finally, we should note that in our analysis, a number of events of interest might occur which are expected to have different relationships with appropriate covariates. Namely, for each company, we're interested in observing several events of interest: obtaining next round of financing, merger or acquisition and going public through IPO. While we expect obtaining next round of financing to be positively related to performance metrics such as Òsocial feedbackÓ data, we expect M\&A exists to be much independent or perhaps even negatively related to company performance.

In order to address this issue, we formulate model in the context of competing risk, corresponding to scenario where events might terminate due to more than one event (obtaining next round of financing, exit via M\&A or IPO). In estimating such model, we take the simplistic latent or cause-specific approach by introducing assumption of independence of competing risks. This assumption is somewhat justified in our case given that the data is transformed in the way that all events of interest are ÒterminalÓ and that exit options of companies generally tend to be independent and only related to firm-specific factors. In this model, we assume that there are $K$ specific outcomes or destination states and it is assumed that there exists a potential or latent failure time associated with each outcome (for example Ð likelihood of obtaining next round of financing vs. likelihood of being acquired). For $K$ possible outcomes, there are $T_k$ possible duration times, but we only observe the shortest time $T_k = min\{T_1 ... T_k\} = T_c$, where $T_c$ represents the duration time associated with observed "cause" of event.

Main idea of latent approach is the fact that if there are $k$ possible outcome states, the overall survivor function can be partitioned into ÒmarginalÓ survivor function, each corresponding to one of $k$ possible destination states. Assuming that there are $n$ observations, individual contribution of event type $k$ occurring at observation $I$ is given by:

\begin{center}
$L_i = f_k(t_i | X_{ik}, \beta_k) \prod_{k \neq r} S_r (t_i | X_{ir}, \beta_r)$
\end{center}

where $r$ represents product term implying that the product is taken over all states except $k$.

Likelihood function for full sample can then be represented in terms of number of observations occurring for each of $K$ outcomes:

\begin{center}
$L = \prod_{k=1}^{r} \prod_{i=1}^{n_k} f_k(t_i | X_{ik}, \beta_k)^{\delta_{ik}} S_k (t_i | X_{ik}, \beta_j)^{1 - \delta_{ik}}$
\end{center}

where $\delta_{ik}$ represents censoring indicator given as:

\begin{center}
$\delta_{ik} = \begin{cases} 1, & \mbox{if } i\mbox{ occurred due to k} \\ 0, & \mbox{otherwise} \end{cases}$
\end{center}

In this way, overall likelihood function is factored into $k$ sub-contributions where failures caused by risks other by k are treated as right-censored. In our analysis, we estimate competing risk model by considering two possible events of interest: obtaining next round of financing and exit through M\&A (number of IPO exit entries are too small to allow for detailed analysis). For each of possible event of interest, we estimate separate Cox proportional hazard model in which we consider events as right-censored if the type of competing event does not correspond to type of competing event being analyzed in given model.

In given context, our main approach in testing of given set of hypothesis of interest is based on estimation of Cox model of the form:

\begin{center}
$h_i(t) = h_0(t)e^{(\beta_1x_{i1} + ... + \beta_k x_{ik})}$
\end{center}

with covariates corresponding "social feedback" variables, and a number of "control" variables suggested by literature and dependent variable corresponding to the duration between two consecutive investment rounds (or censoring in case that next round of financing has not occurred). In order to test for potential violation of proportional hazard assumptions we use the method of Schonfeld residuals and make necessary adjustments to the model in order to eliminate time-dependent effects. We use likelihood-ratio and Wald test in order to assess the overall significance of obtained model. In case that obtained model is shown to be significant, we test for hypothesis of Òsocial feedbackÓ impact to increase likelihood of obtaining next round of financing and shortening of duration between financing rounds by testing for significance and sign of Òsocial feedbackÓ variables. In case that given factors are shown to be insignificant, both hypothesis are rejected. Otherwise, significance of given coefficients $\beta_j$ and values of $(e^{\beta_j} - 1) * 100$ greater than 1, should provide support for acceptance of given hypothesis.

In order to test for hypothesis of independence of M\&A exits of "social feedback" data, we perform the estimation of proposed competing-risk model and analyze the model of M\&A exists in which all other event types are treated as right-censored. We test for overall significance of this model and significance and sign of individual "social feedback" variables. Obtaining model that has overall significance, particularly in "control" variables, but in which "social feedback" variables are either not significant or have a sign that indicates a relatively weak relationship, should provide a support for the hypothesis of relative independence of M\&A exit likelihood of company's consumer-centric performance.

\chapter{Data Collection}

In order to carry out the desired analysis, we compile a novel dataset consisting of detailed information about VC-funded Internet Technology companies and associated data regarding Òsocial feedbackÓ context of financing, including website traffic and search trends data. In aggregating the dataset on Venture Capital investments, we focus on publicly available sources of information and consult multiple sources in order to ensure quality and accuracy of final data. We focus on two main sources of information - VentureDeal and CrunchBase databases, but also consult a number of secondary sources such as PWC MoneyTree and LexisNexis in order to resolve any inconsistencies that might occur in data aggregation process.

VentureDeal \textit{(http://www.venturedeal.com/)} database represents comprehensive source of daily updated information on US-based venture-backed technology companies, venture capital firms, senior management, company financing and M\&A transactions, based on public domain data. As of July 2012, database contains detailed information on 13,673 companies having received at least one round of VC financing, 2,023 active Venture Capital Investors and total of 25,921 investment transactions. Out of this, there are total of 2,315 companies from Internet Technology sector with 4095 associated investment rounds and 827 investors. Information about venture-funded transactions in the database covers the time interval dating from January 1 2003 to present. Database content is updated daily with each new transaction public announcement and information about existing transactions is periodically revisited in order to reflect any new public information unavailable at the time of entry (such as information being exposed through SEC filling data etc.)

CrunchBase \textit{(http://www.crunchbase.com/)} represents free and open repository of technology companies, people and investors with particular focus on high-tech sectors such as IT, Internet and Biotechnology. CrunchBase is developed and maintained by TechCrunch, the most influential technology blog in United States. Unlike VentureDeal, which is centrally curated and based only on official transaction announcement, CrunchBase is collaboratively maintained by the community of technology professionals. Each member can contribute knowledge to the repository and while all of updates go through an approval process before being made available online, actual approval mechanisms uses a much wider range of information sources than traditional Venture Capital databases. This makes given data much more comprehensive, especially in the case of early-stage investments and companies that haven't received initial VC funding. In these areas, CrunchBase information can even be superior than the one provided by commercial VC databases \cite{WB13}. As of July 1st 2012, the database of CrunchBase included information about 95,284 companies, 8,013 financial organizations and 29,583 funding rounds and is growing at the pace of more than 5,000 new entries and 12,000 updates on average per month. It offers almost complete coverage of start-ups and investors in the Internet sector, including the relationships between them. Companies in CrunchBase database are separated into 18 different categories including Consumer Web, Software, eCommerce, Search and others. CrunchBase dataset has first been introduced in literature by \cite{BS09} in analysis of effects of financial crisis on venture capital investments. Since then, a number of publications have used this data in order to analyze various aspects of VC investments, such as role of social capital in startup-funding \cite{AEA10}, impact of co-investment networks on start-up performance \cite{WB13} and influence of geographical proximity and industry similarity on investment choices \cite{DO10}, \cite{BER11}.

In order to assess relative quality and accuracy of information in VentureDeal and CrunchBase databases, we perform spot check of Internet technology companies from both databases and find that CrunchBase provides much more extensive information about seed and early-stage angel investments whereas VentureDeal provides higher-quality information about later stage financing (Series A and beyond). This is somewhat expected given the fact that VentureDeal database is based only on official investment announcements. Given the the more "formal" nature of VentureDeal database we use it as a primary source of funding information and use Crunchbase as a source of information about early stage rounds and verification of financing data. In creation of Venture Capital Investment dataset user in this research, we obtain the complete set of 2,315 US companies from ÒInternetÓ industry category in VentureDeal database. For each company we obtain status, description and company website as well and detailed data about each financing round, acquisition or IPO including information release dates, transaction types, amounts and participating investors. For each entry we consult CrunchBase database in order to validate given entry and fill out any missing information as well as detect potential investment round data not available in VentureDeal database. In case of inconsistencies between VentureDeal and CrunchBase entries, we consult external public sources using LexisNexis service in order to find relevant public information in support of either of entries.

In addition to Venture Capital investment data, we also need to obtain information about "social feedback" on the Web relevant to start-up companies in given data set. In order to do so, we use two additional sources of data - Alexa Web Information Service and Google Trends. 

Alexa Web Information Service \textit{(http://www.alexa.com/)} collects and aggregates historical traffic data from millions of Alexa Toolbar users and other, diverse traffic data sources. For each website, Alexa provides historical measurements of daily reach, number of page views and traffic rank. Daily reach represents estimated percentage of all Internet users who visit a given website in certain time interval. For example, daily reach of 0.08 for website \textit{airbnb.com} means that out of all global Internet users, estimated 0.08\% of them visited \textit{airbnb.com}. Alexa's one-week and three-month average reach are computed by averaging daily measurements over specified time period. Page Views represents measure of the number of unique web pages viewed by individual site visitor. This figure is reported as page views per user number, which represents average number of unique pages viewed per user in given time interval. Finally, Traffic Rank metrics for each website is derived by combining page views and reach metrics for specified time period by averaging appropriate daily measurements. This metrics provides a convenient way of measuring the relative "impact" of every individual website on the Internet. Based on this value, Alexa assigns each individual site a single "Alexa Traffic Rank" figure and publishes regularly updated list of "Top SitesÓ \textit{(http://www.alexa.com/topsites)} according to their rank.\\

\begin{figure}[htb]
\centering
\includegraphics[width=0.8\textwidth]{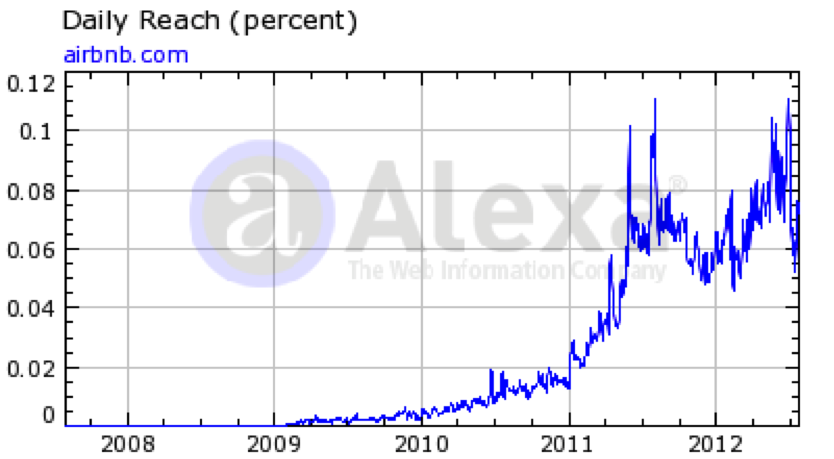}
\caption{Example Alexa Traffic Data Entry}
\end{figure}

For the purpose of our research, we compile the dataset consisting of historical time series of Alexa traffic measurements for each Internet technology company in our Venture Capital investment dataset. In order retrieve this data, we use Alexa Web Information Service API (\textit{http://aws.amazon.com/awis/}) which provides a way of programmatic access to historical Alexa traffic measurements ranging from August 1 2007 to present. Using a specially created application, for each individual company, we retrieve it's website (or website of it's primary product) and use Alexa API in order to retrieve historical values of daily reach, page views and traffic rank for the period from August 1 2007 to August 1 2012.

Google Trends (http://www.google.com/trends/) represents a service provided by Google, enabling access to information about the number of web searches that have been undertaken for a particular search term, relative to the overall number of searches completed by Google Search Engine within certain time period. The "query share" defined in this manner is intended to represent the user's propensity to search for a certain topic on Google on a relative basis. In order to achieve this, maximum query share is normalized to be 100 and query share at initial date being examined is normalized to zero. Relative search volumes are aggregated weekly in the form of \textit{search volume index} along with related \textit{news reference volume}, representing number of times associated topics appears in Google News stories. The actual queries are determined using "broad matching", which means that multiple queries with the similar meaning might be accounted as the same search term. Due to privacy considerations, data is computed using sampling method and is only tracked for terms for which there are meaningful search volumes. For each sample entry, information about country, city and language is recorded and provided in the form of aggregate counts along with the associated traffic data. Historical data in Google Trends database contains entries ranging from January 1 2004 till present, aggregated on weekly basis. Google Trends data has been extensively used in literature in wide variety of applications such as determining social interest in health issues \cite{BO10}, detecting disease outbreak \cite{CM09}, short-term of forecasting of economic indicators \cite{CV09A}, forecasting hotel room demand \cite{PAN12} and predicting move box-office revenue \cite{GOEA10}.

\begin{figure}[htb]
\centering
\includegraphics[width=0.8\textwidth]{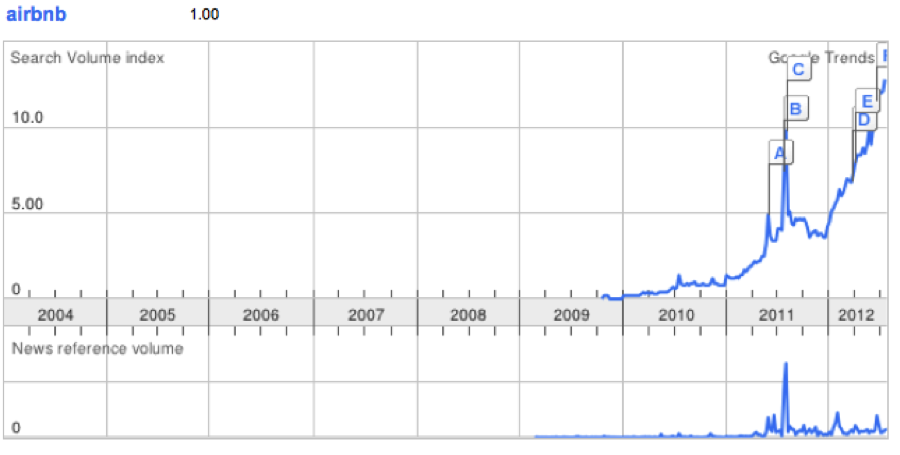}
\caption{Example Google Trends Entry}
\end{figure}

In order to construct dataset of interest, for each company in our Venture Capital dataset, we define a single search term that is most likely to correspond to company name or it's primary product. For each such entry, we query Google Trends and determine whether meaningful search volume exists in order for Google to aggregate appropriate search volume data. In case such information exists, we use Google Trends filter function to further restrict results to queries originating from United States. If this still results in meaningful search volumes, we use Google Trends CSV export function in order to retrieve appropriate time series data. In case that there is not enough search volume data for Google to show appropriate trends, we associate given company with "no visible search volume trend" entry.

\chapter{Dataset Description}

Based on methodology described in previous chapter, we create final dataset that we will use in addressing the research questions at hand. Dataset consists of entries corresponding to duration of time between two events of interest observed in the period between January 1 2004 and August 1 2012 and appropriate set of covariates. Each individual observation entry contains the following fields:

\begin{itemize}
  \item \textit{Company Name} - Name of the company for which given entry is observed.
  \item \textit{Company Type} - Type of Internet Technology company in given observation. All companies are classified into three categories: Consumer Product, Enterprise Product and Platform, with Consumer Product corresponding to companies providing online applications targeted at broad set of consumers (Facebook, Twitter...), Enterprise Product corresponding to specific applications targeted at enterprises (PBWorks, Zendesk...) and Platform corresponding to advertising and lead generation platforms (Adometry, Invite Media...)
  \item \textit{Investment Type} - Type of event that has occurred in a given entry. Valid values include: Venture Equity, M\&A, IPO and no event. Venture Equity corresponds to obtaining next round of VC financing, M\&A corresponds to company's exit via being acquired or merged with another company and IPO corresponds to companyÕs exit via initial public offering. Finally, "no event" correspond to case in which no financing or liquidity event for given company has occurred within observed time interval.

\item \textit{Investment Amount} - Financing amount that has been received as a result of a given event, or zero in case of entries for which no event has occurred or proceeding details have not been disclosed

\item \textit{Total Capital Raised} - Total capital that given company has raised in all previous rounds, not accounting for any potential proceedings from financing occurred in given entry. For companies for which historical data is available prior to January 1 2004, any previous financing information is taken into account when computing this variable.

\item \textit{Round Name} - Descriptive name of the round of financing obtained in given entry (Seed, Series A, Series B...)

\item \textit{Round Number} - Integer representing order of occurrence of given round in complete history of investments for given company. For companies with available history prior to January 1 2004, all older investments are taken into account when computing this variable.

\item \textit{Weeks Since First Investment} - Number of months that have elapsed since given company has received first financing round, computed taking into account all available historical information including any known funding rounds that have occurred prior to January 1 2004.

\item \textit{Weeks Since Last Investment} - Number of weeks elapsed since given company has received the last round of financing. Similar as with previous duration variables, any known financing events that have occurred before the start of the observed interval are taken into account. This variable represent main dependent variable that we aim at modeling in the proposed Cox Proportional Hazard model.

\item \textit{Event Has Occurred} -  Binary variable representing censoring variable in proposed Cox model with value 1 in the case when event of interest had occurred and value of 0 when event did not occur (either due to the fact that given company has failed to exit or attract next round investments or due to right-censoring of observations at the end of observed interval)

\item \textit{Has Trends Data} -  Binary variable indicating whether Google Trends information exists for a given company, with value 1 in cases where trends data exist and 0 in cases where it doesn't. We should note that although primary reason for nonexistence of trends data would be lack of brandÕs prominence in social feedback reflected via Google search, there are several cases in which brand name is a common dictionary keyword (Lemon, Science, Rocket...). Given that in such cases it is impossible to distinguish company-specific from general search trends, such entries are marked as no trends data.

\item \textit{Trends Delta} - Percentage change in Google Search Trends data within the interval for which given entry has been observed. This value should be present for entries for which trends data exists.

\item \textit{Has Traffic Data} - Binary variable indicating whether Website Traffic information is available for given company within observed time interval. While Website Traffic information is generally available across all companies in the dataset, due to the limitation of our traffic information sources, it is only present for the entries in the period between August 1 2007 and August 1 2012.

\item \textit{Traffic Delta} -  Percentage change in company's website traffic within the interval for which given entry has been observed. This value should be present for all entries for which traffic data exists.

\end{itemize}

We should also note that, given the fact that this dataset is defined in the context of duration of time between investments, all companies that have received only a single financing round will have only a single right-censored entry. Additionally, all companies that contain exit entries (M\&A and IPO), will not contain any censored entries, as is it assumed that given companies will not require any additional VC funding and hence should be removed from further consideration in the dataset.

\chapter{Data Analysis}

Final dataset consists of 7453 entries, corresponding to total of 2048 companies, with each entry representing duration of time between two consecutive financing or liquidity events for single company and appropriate set of covariates. Survival times between each two events for given company are represented via "Weeks Since Last Investment" variable while right-censored observations for which next-stage event has not occurred yet are marked via "Event Has Occurred" variable. Dataset covers period from Jan 1 2004 to August 1 2012. Summary of different exit types available in the data is given in Table 6.1.\\

\begin{table}[H]\centering
\begin{tabular}{@{}llllr@{}} \toprule
Venture Equity & M\&A & IPO & No Event & Total\\ \midrule
2717 & 317 & 16 & 1703 & 4753\\
(57.16\%) & (6.69\%) & (0.34\%) & (35.83\%) & \\ \bottomrule
\end{tabular}
\caption{Event types present in the dataset}
\end{table}

We note that in given dataset, IPO exists are certainly underexpressed, primary due to the fact that we're only observing new investment rounds in given period and that the IPO market for Internet companies has been somewhat unfavorable in the period after the dot-com boom and only started to recover after 2010. On the other hand, we notice that there are substantial number of M\&A exists in the dataset, which seem to be a primary exit route for Internet companies in given time period. We should also note that roughly 35\% of entries are right-censored, representing companies that still haven't exited or received next round of financing at the end of observed interval.

Total investment amounts received by companies in the dataset for different event types are represented in Table 6.2.

\begin{table}[H]\centering
\begin{tabular}{@{}llllllr@{}} \toprule
& Min & 1st Qtl. & Median & Mean & 3rd Qtl. & Max\\ \midrule
Venture Equity & 0.018 & 1.5 & 4.2 & 10.28 & 10.0 & 1500.0\\
M\&A & 0.5 & 13.35 & 60.0 & 174.40 & 179.50 & 2147.0\\ 
IPO & 54.17 & 75.73 & 110.8 & 1160.0 & 220.80 & 16000.0\\
Total & 0.018 & 1.250 & 4.000 & 16.810 & 9.000 & 16000.0\\ \bottomrule
\end{tabular}
\caption{Summary of Investment Amount Distribution (in millions \$)}
\end{table}

Note that although most investments tend do be concentrated in certain range specific to particular investment type, given a small number of very successful companies, overall range of investment amounts tend to vary in the order of magnitude. Therefore, in our analysis we use logarithm of investment amounts rather than actual values. 

Distribution of round numbers across all entries in the dataset is represented using the histogram in Fig 6.1 and associated summary is given in Table 6.3.

\begin{figure}[htb]
\centering
\includegraphics[width=0.75\textwidth]{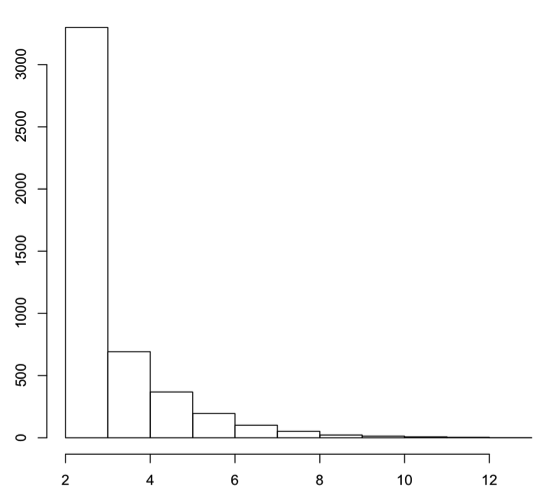}
\caption{Histogram of round numbers for all entries in the dataset}
\end{figure}

\begin{table}[H]\centering
\begin{tabular}{@{}lllllr@{}} \toprule
Min & 1st Qtl. & Median & Mean & 3rd Qtl. & Max\\ \midrule
2 & 2 & 3 & 3.205 & 4 & 13\\ \bottomrule
\end{tabular}
\caption{Summary of Round Number distribution}
\end{table}

From given data, it seems that most of the entries in the dataset tend receive only a first round of financing and fail to obtain subsequent rounds. However, we also note that a small number of very successful startups tend to get financed with large number of rounds, usually in preparation for IPO exit. 

Distribution of company types present in the dataset is given in Table 6.4.

\begin{table}[H]\centering
\begin{tabular}{@{}lllr@{}} \toprule
Consumer Product & Enterprise Product & Platform & Total\\ \midrule
3274 & 1127 & 352 & 4753\\
(68.88\%) & (23.7\%) & (7.4\%) & \\ \bottomrule
\end{tabular}
\caption{Company Types present in the dataset}
\end{table}

In order to illustrate survival profile for all entries in the dataset, we plot a simple Kaplan-Meier nonparametric estimate of $S(t)$, given as:

\begin{center}
$\hat{S{(t)}} = \prod_{t_i < t} { ({{n_i - d_i} \over {n_i}}) }$
\end{center}

where $n_i$ represents number of companies that haven't exited or received financing prior to time $t_i$, while $d_i$ represents number of companies for which event of interest has occurred at given time $t_i$.

\begin{figure}[H]
\centering
\includegraphics[width=0.7\textwidth]{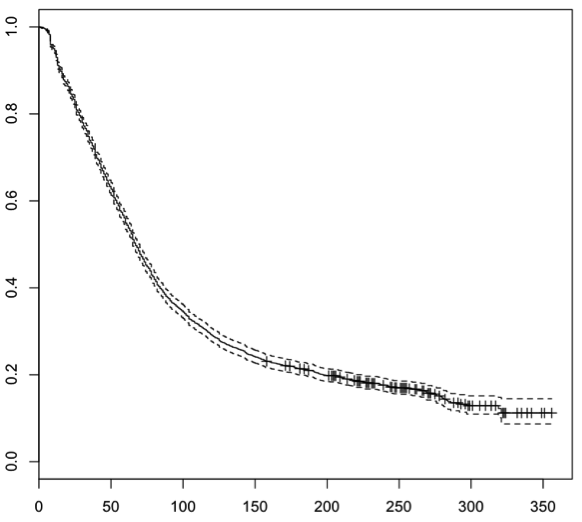}
\caption{ Kaplan-Meier Estimate of survival function for entire dataset}
\end{figure}

\begin{table}[H]\centering
\begin{tabular}{@{}llllr@{}} \toprule
Min & 1st Qtl. & Median & 3rd Qtl. & Max\\ \midrule
1 & 32.2 & 67 & 133.1 & 321 \\ \bottomrule
\end{tabular}
\caption{Summary of times between events (in weeks)}
\end{table}

From Table 6.5, we can conclude that the median time between two subsequent events of interest is 67 weeks (1.28 years).

We also compute Kaplan-Meier estimates of survival function for different investment types (Venture Equity, Debt and IPO), with survival function given in Figure 6.3 and corresponding summary in Table 6.6.

\begin{figure}[H]
\centering
\includegraphics[width=0.7\textwidth]{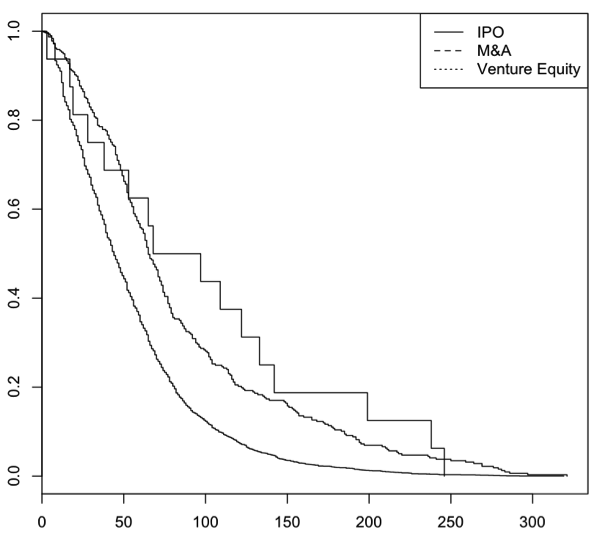}
\caption{ Kaplan-Meier Estimate of survival function for entire dataset}
\end{figure}

\begin{table}[H]\centering
\begin{tabular}{@{}lllllr@{}} \toprule
& Min & 1st Qtl. & Median & 3rd Qtl. & Max\\ \midrule
Venture Equity & 3 & 28 & 82.5 & 133 & 246 \\
M\&A & 2 & 42 & 65 & 104.3 & 297\\ 
IPO & 1 & 22.3 & 44 & 72.1 & 319 \\\bottomrule
\end{tabular}
\caption{Summary of investment type - dependent survival times}
\end{table}

Given results indicate that survival functions for given three types of investment events are different. We also test for this formally using logrank test under the null hypothesis that all groups have identical survival function and reject the hypothesis at the 0.001 significance level.

Finally, we give a brief summary of "social feedback" covariates, represented by Has Trends Data, Trends Delta and Traffic Delta variables. Out of total of 4753 entries, corresponding to 2048 unique Internet companies, total of 1875 (39.4\%) entries has appropriate trends data, corresponding to total of 633 (30.9\%) unique companies.

Summary of Search Trends Delta distribution for all entries in the dataset is given in Table 6.7.

\begin{table}[H]\centering
\begin{tabular}{@{}lllllr@{}} \toprule
Min & 1st Qtl. & Median & Mean & 3rd Qtl. & Max\\ \midrule
-100.0 & 0.0 & 0.0 & 17.38 & 10.44 & 314.20 \\ \bottomrule
\end{tabular}
\caption{Summary of Search Trends Delta Distribution}
\end{table}

We note that given distribution is somewhat positively skewed with mean change in traffic trend corresponding to 10\% increase between two investment rounds. Slight positive bias in search trends distribution is expected given the fact that very existence of search trend assumes pre-existence of positive social feedback trend. Similarly, we note that out of total of 4754 entries, total of 4008 (84.3\%) contain traffic information while 745 (15.7\%) don't.

Traffic delta distribution is somewhat more evenly distributed, as shown in Table 6.8.

\begin{table}[H]\centering
\begin{tabular}{@{}lllllr@{}} \toprule
Min & 1st Qtl. & Median & Mean & 3rd Qtl. & Max\\ \midrule
-100.0 & -72.00 & 0.0 & 43.16 & 49.20 & 505.10 \\ \bottomrule
\end{tabular}
\caption{Summary of Website Traffic Delta Distribution}
\end{table}

This is primary due to the fact that measurements of traffic data are not subject to the type of "thresholding" that is present in the case of Search Volume Trends data.

\chapter{Results}

Based on the methodology outlined so far we proceed with estimation of proposed latent competing risk Cox proportional hazard model. In order to estimate given model we perform separate estimations of models corresponding to particular type of risk with events being treated as right-censored in cases when they don't correspond to the type of risk being estimated. In particular - we estimate two separate models, corresponding to risk of obtaining next round of Venture Capital investments and risk of exiting via M\&A or IPO as well as the "risk-type-oblivious" model corresponding to time to any financing or liquidity event. In each model, we use as covariates a set of "control" variables such as investment amount, elapsed time since initial investment and total raised capital as well as set of variables specific to "social feedback" of interest - presence of trends data and appropriate changes in search volume and website traffic. We also consider company type as a covariate in order to analyze for potential effects of particular product specialization. Results of initial Cox model estimation of a time-to-event data, considering any event (investment or liquidity event) as equal is given in Table 7.1.

\begin{table}\centering
\begin{tabular}{@{}lllllr@{}} \toprule
Covariate name & Beta & Exp(beta) & Se(coef) & Z & $Pr(>|z|)$\\ \midrule
Log(totalCapital) & -0.077599 & 0.925 & 0.01826 & -4.250 & 2.13e-05 \\
roundNumber & 0.109617 & 1.116 & 0.01785 & 6.142 & 8.13e-10 \\
weeksSinceFirst & -0.000932 & 0.990 & 0.00025 & -3.694 & 0.00022 \\
trafficDelta & 0.005317 & 1.005 & 0.00027 & 19.450 & $<$ 2e-16 \\
hasTrendsData & 0.286044 & 1.331 & 0.04051 & 7.061 & 1.65e-12 \\
trendsDelta & -0.002120 & 0.997 & 0.00047 & -4.521 & 6.17e-06 \\
trendsDeltaSign & 0.243903 & 1.276 & 0.05552 & 4.393 & 1.12e-05 \\
companyType=EP & 0.002224 & 1.011 & 0.04481 & 0.050 & 0.9604 \\
companyType=PL & 0.208910 & 1.232 & 0.06970 & 2.997 & 0.0027 \\ 
\end{tabular}

\begin{tabular}{@{}lllllr@{}} \toprule
Concordance= 0.628 (se = 0.006)\\
Rsquare= 0.108\\
Likelihood ratio test = 545.1 on 9 df, p=0\\
Wald test = 551.2 on 9 df, p=0\\
Score (logrank) test = 566.3 on 9 df, p=0~~~~~~~~~~~~~~~~~~~~~~~~~~~~~~~~~~~~~~~~~~~~\\ \bottomrule
\end{tabular}

\caption{Initial estimate of risk-oblivious Cox model}
\end{table}

Based on likelihood ratio, Wald and logrank tests, we reject the null hypothesis that all of the coefficients $\beta$ are zero. Therefore, given model seems highly significant with $p$ values of appropriate tests close to zero. Additionally, individual parameter tests indicate that all of the variables in the model are significant at 0.001 level, except for "Enterprise Product" value of categorical variable \textit{companyType}. However, based on Grambsch and Therneau test of Schoenfeld residuals, we're not able to reject the null hypothesis of time independence for \textit{roundNumber} and \textit{weeksSinceFirstInvestment} variables, which can be interpreted as strong evidence that given variables have non-proportional hazards. In order to address this, we reformulate given model by introducing interaction effects between given covariates and time. In particular, we introduce two interaction terms to the model: interaction between \textit{roundNumber} and \textit{yearsSinceFirstInvestment} and interaction between \textit{weeksSinceFirstInvestment} and \textit{weeksSinceLastRound} variables. Obtained model is represented in Table 7.2. Based on likelihood ratio, Wald and logrank tests, we can't reject hypotnesis that all coefficients $\beta$ are zero, which indicates the significance of obtained model. We also note that all individual parameter estimates except for EP value of categorical variable \textit{companyType} are significant at 0.001 level, though beta coefficients somewhat differ from estimates obtained in original model.\\

\begin{table}\centering
\begin{tabular}{@{}lllllr@{}} \toprule
Covariate name & Beta & Exp(beta) & Se(coef) & Z & $Pr(>|z|)$\\ \midrule
Log(totalCapital) & -0.071607 & 0.931 & 0.0188 & -3.804 & 1.4e-04 \\
roundNumber & 0.140348 & 1.151 & 0.0256 & 5.4733 & 4.4e-08 \\
weeksSinceFirst & 0.015403 & 1.016 & 0.0006 & 24.848 & $<$ 2e-16\\
trafficDelta & 0.005174 & 1.005 & 0.0003 & 18.783 & $<$ 2e-16 \\
hasTrendsData & 0.291102 & 1.338 & 0.0407 & 7.1584 & 8.2e-13 \\
trendsDelta & -0.001446 & 0.999 & 0.0014 & -3.179 & 1.5e-03 \\
trendsDeltaSign & 0.145077 & 1.156 & 0.0531 & 2.7353 & 0.0062 \\
companyType=EP & 0.000453 & 1.000 & 0.0457 & 0.0101 & 0.99 \\
companyType=PL & 0.189269 & 1.208 & 0.0701 & 2.703 & 0.0069 \\ 
roundNumber:\\ 
~yearsSinceFirst & -0.073076 & 0.930 & 0.0065 & -11.22 & $<$ 2e-16 \\ 
weeksSinceFirst:\\ 
~weeksSinceLast & -0.000109 & 1.000 & 3.6e-06 & -30.21 & $<$ 2e-16 \\ 
\end{tabular}

\begin{tabular}{@{}lllllr@{}} \toprule
Concordance= 0.739 (se = 0.006)\\
Rsquare= 0.319\\
Likelihood ratio test = 1828 on 11 df, p=0\\
Wald test = 1779 on 11 df, p=0\\
Score (logrank) test = 1434 on 11 df, p=0~~~~~~~~~~~~~~~~~~~~~~~~~~~~~~~~~~~~~~~~~~~~\\ \bottomrule
\end{tabular}

\caption{Risk-oblivious Cox model with added interaction terms}
\end{table}

By interpreting the values of significant coefficients in the model we can derive the following conclusions:\\

\textit{Control variables}

\begin{itemize}
\item Increase in total amount of capital raised is associated with decreased hazard of occurrence of the event of interest. Companies which manage to raise larger total amounts of capital tend to prolong duration between a given round and subsequent round of financing or liquidity event. This can be justified by the fact that larger amounts of raised capital provide companies with longer time windows in which next round of capital infusion will not be required
\item Increase in number of received financing rounds is associated with increased hazard of event of interest. That is - companies that have received more rounds of financing are more likely to obtain future financing as well, and with shorter duration. Obtained coefficient indicates that each new round of financing increases likelihood of exit or receiving a follow up financing by 15 per cent.
\item Increase in elapsed time since initial investment is associated with increased hazard of event of interest. Obtained coefficient indicates that with each elapsed year since initial round of funding, likelihood of exit or receiving new financing round increases by 83 per cent (derived from obtained weekly estimate).
\item Internet companies operating in the area of advertising and lead generation platforms are 28\% more likely to obtain next round of financing than companies developing consumer or enterprise oriented products.\\
\end{itemize}

\textit{Social feedback variables}

\begin{itemize}
\item Presence of brand name in Google Search Trends data is associated with increased hazard of event of interest. In particular Ð companies for which a trend entry exists in publicly available web search data are 33.8\% more likely to obtain next round of financing or exit via merger or IPO than companies for which no such trend exists.
\item Increase in search volume over a given time period is associated with increased hazard of event of interest, leading to 15.6 per cent increase in likelihood of exiting or obtaining financing. Equivalently, decrease in search volume over the same period will lead to 15.6 decrease in hazard rate of event of interest
\item Positive percentage change in website traffic volume for given company is associated with increased hazard of event of interest. A 100 per cent increase in website traffic volume will lead to 50 per cent increase of likelihood of exiting or obtaining next financing round. Similarly, a 100 per cent decrease in website traffic volume will lead to 50 per cent decrease of hazard rate of same events.
\end{itemize}

Therefore, we can conclude that given results provide strong support for hypothesis that positive trends in "social feedback" data are expected to increase likelihood of next round of financing and decrease duration between investments for technology companies (corresponding to hypothesis H1 and H2). In addition to given model, we also estimate a latent competing risk model by separately considering risk of receiving next round of financing and exit via merger or IPO. Results of Cox model estimation for receiving next round of VC investments with M\&A and IPO events treated as right-censored are given in Table 7.3.

\begin{table}\centering
\begin{tabular}{@{}lllllr@{}} \toprule
Covariate name & Beta & Exp(beta) & Se(coef) & Z & $Pr(>|z|)$\\ \midrule
Log(totalCapital) & -0.082420 & 0.921 & 0.0201 & -4.110 & 3.96e-05 \\
roundNumber & 0.177812 & 1.195 & 0.0276 & 6.438 & 1.21e-10 \\
weeksSinceFirst & 0.017696 & 1.018 & 0.0007 & 24.787 & $<$ 2e-16\\
trafficDelta & 0.005490 & 1.006 & 0.0003 & 18.811 & $<$ 2.6e-13 \\
hasTrendsData & 0.314539 & 1.370 & 0.0430 & 7.313 & 2.6e-13 \\
trendsDelta & -0.001531 & 0.998 & 0.0005 & -3.115 & 0.00184 \\
trendsDeltaSign & 0.120114 & 1.128 & 0.0567 & 2.120 & 0.03403 \\
companyType=EP & 0.014759 & 1.015 & 0.0477 & 0.309 & 0.75717 \\
companyType=PL & 0.262957 & 1.301 & 0.0730 & 3.604 & 0.00031 \\ 
roundNumber:\\ 
~yearsSinceFirst & -0.094553 & 0.910 & 0.0077 & -12.290 & $<$ 2e-16 \\ 
weeksSinceFirst:\\ 
~weeksSinceLast & -0.000138 & 1.000 & 4.5e-06 & -30.687 & $<$ 2e-16 \\ 
\end{tabular}

\begin{tabular}{@{}lllllr@{}} \toprule
Concordance= 0.758 (se = 0.006)\\
Rsquare= 0.334\\
Likelihood ratio test = 1930 on 11 df, p=0\\
Wald test = 1734 on 11 df, p=0\\
Score (logrank) test = 1331 on 11 df, p=0~~~~~~~~~~~~~~~~~~~~~~~~~~~~~~~~~~~~~~~~~~~~\\ \bottomrule
\end{tabular}

\caption{Cox model for risk of obtaining next round of VC financing}
\end{table}

Obtained model is highly significant and all obtained coefficient preserve the same signs as in the "common risk" model. Certain parameter estimates have slightly different coefficients indicating somewhat stronger relationships than in the common model. For example, in the case of venture capital investment risk, presence of brand in the Google Search Trends data is associated with 37 per cent increase in investment hazard risk as opposed to 33 per cent in the common model. Therefore, we can conclude that obtained results provide support for hypothesis H1 and H2, similar to the case of "common risk" model.

Finally, we estimate the Cox model of "exit risk" corresponding to exit via merger or acquisition, with venture capital investment entries treated as right-censored. Results of given estimation are described in Table 7.4. Based on likelihood ratio, Wald and logrank test, we conclude that given model is still highly significant, but unlike in the case of "common" and venture capital investment risk, we can't establish a significant relationship between duration of time-to-exit and given "social feedback" covariates. On the other hand, we see that round number and number of weeks elapsed since first investment are still associated with increased hazard of exit and that obtained coefficients are similar as in the "common" model, indicating that company's age and number of received investment rounds significantly increase companyÕs likelihood of exit via M\&A.

\begin{table}\centering
\begin{tabular}{@{}lllllr@{}} \toprule
Covariate name & Beta & Exp(beta) & Se(coef) & Z & $Pr(>|z|)$\\ \midrule
Log(totalCapital) & 0.0324 & 1.033 & 0.0557 & 0.5813 & 0.56 \\
roundNumber & 0.232 & 1.261 & 0.0771 & 3.0061 & 0.0026 \\
weeksSinceFirst & 0.0176 & 1.018 & 0.0015 & 11.7537 & $<$ 2e-16\\
trafficDelta & 0.00237 & 1.002 & 0.0009 & 2.7306 & 0.00632 \\
hasTrendsData & -0.0121 & 0.988 & 0.129 & -0.0939 & 0.92518 \\
trendsDelta & -0.00075 & 0.999 & 0.0013 & -0.5684 & 0.56976 \\
trendsDeltaSign & 0.181 & 1.198 & 0.161 & 1.1262 & 0.26006 \\
companyType=EP & -0.146 & 0.865 & 0.138 & -1.0529 & 0.29237 \\
companyType=PL & -0.453 & 0.636 & 0.264 & -1.7159 & 0.08618 \\ 
roundNumber:\\ 
~yearsSinceFirst & -0.0856 & 0.918 & 0.0163 & -5.2630 & $< $1.42e-07\\ 
weeksSinceFirst:\\ 
~weeksSinceLast & $-5.9e-05$ & 1.000 & $6.0e-06$ & -9.6767 & $<$ 2e-16 \\ 
\end{tabular}

\begin{tabular}{@{}lllllr@{}} \toprule
Concordance= 0.718 (se = 0.019)\\
Rsquare= 0.047\\
Likelihood ratio test = 228.3 on 11 df, p=0\\
Wald test = 313.5 on 11 df, p=0\\
Score (logrank) test = 292.8 on 11 df, p=0~~~~~~~~~~~~~~~~~~~~~~~~~~~~~~~~~~~~~~~~~~~~\\ \bottomrule
\end{tabular}

\caption{Cox model of Òliquidity eventÓ risk}
\end{table}

This result provides a strong support for hypothesis H3, that M\&A exits of VC-funded technology companies are not likely to be significantly determined by trends in "social feedback" data. This seems to suggest that M\&A exits of Internet companies are not necessary related to company's consumer-oriented performance. One possible explanation for this might be that most of exits in given dataset represent either talent or technology acquisition and therefore are not significantly related to company's user base or it's business growth, for which proposed "social feedback" data should represents a convenient proxy. Additionally, such result might be interpreted as support for VC's ability to orchestrate successful exits in cases when company develops great team or technology portfolio, but fails to generate high-growth consumer business around it.

\chapter{Summary}

In this research, we have examined the relationship between staging of Venture Capital investments and available exit options for Internet companies as a function of company's performance, measured using proxy in the form of publicly available "social feedback" data such as Web search trends and website traffic information. Primary objective of this research was providing support for "learning hypothesis" regarding Venture Capitalist's decisions to stage capital infusions in new startup companies. We find support for the hypothesis that VCs actively use performance-related information in order to evaluate prospects of new ventures throughout their development and allocate future investments in the way that maximizes overall expected return on the portfolio. The particular significance of this result is that it suggests that staging of investments is not a mere technical matter and provides an explanation for the empirical fact that large number of Internet and Technology startups in general fail to receive future funding early in its development. We also show results in support of VCs ability to orchestrate exits even in the scenarios where new companies fail to develop successful business around potentially valuable technology or great team. Such result is of particular importance for entrepreneurs as it suggests that Venture Capital is likely to reduce risk of new project failure, especially in the case of ventures with unknown prospects. Therefore, such means of funding might be much more attractive than alternatives such as debt financing or "bootstrapping".

In order to obtain given results, we have compiled a unique dataset consisting of publicly available data on VC financing of Internet startups in the post dot-com period and associated longitudinal data on certain "social feedback" variables for periods between each two financing rounds. By applying methods of survival analysis to given data, we have managed to establish strong relation between given publicly-available indicators, representing a proxy of consumer interest and adoption of new product, and VCs perception of project quality, reflected in decision making regarding projectÕs future financing. Particular importance of this aspect of research lies in the fact that given analysis has been conducted exclusively focusing on publicly available data and therefore provides an objective and reproducible way of gaining insights on prospects of new ventures and performing industry-level analysis. To the best of our knowledge, this is the first publication that leverages given indicators in gaining insights into development prospects of private companies and VC decision-making. A particularly interesting implication of given research might be the ability to construct an "index" of prospects of new technology ventures, measuring likelihood of success of new companies in different market segments relative to invested amount of capital. This view of "innovation capital at risk" might be of particular interest for policy makers of industry-level analysts.

We should also acknowledge that there are certain limits to generalizability of obtained results. Namely, current publication primary focuses on analysis of Internet companies, which are somewhat specific, particularly in ability to launch products early in company's lifecycle and continue business development for a long time without reaching profitability and therefore depending on future Venture Capital investments. Such practices are not necessary the rule in certain industries such as Semiconductors or Biotechnology, and therefore applicability of obtained results is likely to be limited to narrow set of similar industries. Additionally, we should note that obtained models are limited to variables that are publicly available across all of the companies in the industry. 

As indicated by previous research, a number of other firm-specific variables exist which are shown to contribute to VC decision-making. Obtaining such variables for all companies in single industry would be a significant challenge, but a potential follow-up research could be conducted focusing on a smaller sample of companies and attempting at obtaining this information by means of an interview of individual companies from the sample. Performing such analysis on a model given in this research might provide a way to support findings of the research, in light of full information about all aspects of company's development, both public and private. Another aspect of this research that requires additional validation, would be repeating given analysis by using VC investment data obtained from commercial private equity databases such as ThomsonOne and DowJones VentureSource. While best effort has been applied in making sure that obtained VC investment data is of highest quality, this research is primary based on public data and it's validation against commercial data sources should be critical, especially given the inconsistencies occurring even between such commercial sources \cite{MAEA11}. Usage of these data sources should also be critical for extending given analysis to the case of different industries, for which publicly available financing data tends to be scarcely available.

An important topic for future research would be extension of the notion of "social feedback" to indicators derived from additional publicly available "social media" sources such as Twitter, Facebook, Digg and others. As noted in literature review, a number of research publications have already established the usability of such sources in expressing consumer sentiment towards brands and products. Such data should provide a way of extending given results into segments other than "dot-com" companies, and might reflect consumer opinions in much more detail, including positive or negative sentiment and potential relationships to competing products or brands. Such findings would provide even more accurate proxy for consumer demand and new ventureÕs prospects, and should provide us with a way of obtaining additional insights into the process of VC decision making.

\end{document}